\documentstyle[twocolumn,aps]{revtex}
\draft
\begin{document}
\title{ Quantum Hall Ferromagnets}
\author{Tin-Lun Ho}
\address{
\cite{per}Physics Department,  The Ohio State University, Columbus, Ohio
43210\\
National High Magnetic Field Laboratory,  Florida State
University, Tallahassee, FL 32306-4005 \\
\vspace{.3in}\begin{minipage}[t]{5.5in}  \rm
It is pointed out recently\cite{Girvin}
that the $\nu=1/m$ quantum Hall states in
bilayer systems behave like easy plane quantum ferromagnets.
We study the magnetotransport of these systems using their
``ferromagnetic" properties and a novel spin-charge relation of their
excitations.
The general transport is a combination of the
ususal Hall transport and a time dependent transport with
$quantized$ time average. The latter is due to a phase slippage
process in $spacetime$ and is characterized by two topological
constants.
\end{minipage}}
\maketitle

\bigskip

Recent technological advances in producing high mobility bilayers
have provided opportunities for studying quantum Hall (QH)
systems with
internal degrees of freedom\cite{expt}. In this case, the internal degrees
of freedom is a
pseudo-spin 1/2 labelling the upper and lower layer.
There are experimental evidence\cite{ATT} indicating that under
appropriate conditions, QH states (at $\nu=1/2$ and
$\nu=1$) can be stabilized by correlations
between layers and have little to do with tunneling
between them. It is conceivable that similar correlated states
will be found at lower filling factors. Bilayer QH states
with $\nu=1/m$, ($m$ odd) are believed to be the
so-called $(mmm)$ state. It is realized recently that they are very
unusual QH fluids.
Wen and Zee\cite{WenZee} showed that these states can produce
Josephson like tunneling between the layers. Very recently, a
number of authors\cite{Girvin} have pointed out that these states behave like
easy plane quantum ferromagnets and have studied
possible phase transitions in these systems.

The purpose of this paper is to study the magnetotransport
of QH ferromagnets (QHFs).
The difficulty of the study arises from the fact that the usual
procedure of extracting conductivity tensor
(say, using flux insertion arguments or
Chern-Simon type of theories) breaks down when applied to the $(mmm)$
state\cite{comment1}.
Here, we take a new approach which amounts to studying the
hydrodynamics of QHFs.
Our approach recovers the known results for ordinary QH fluids,
and reveals a new transport mode in QHF.
A general QHF transport is a combination of usual QH
trasnport and a genuinely time dependent transport with
{\em quantized} time
average.  The latter corresponds to a phase slippage process
which turns out to be a continuous
nucleation of {\em coreless vortices} in {\em spacetime}.
The conducitivity tensor for the time average currents
is characterized by two
topological constants specifying the nucleation
process.
It is interesting to note that there is a close
analogy between QHF and
superfluid $^{3}$He-A. They have similar order parameters,
topological excitations, and vortex
nucleation processes.

To determine the magnetotransport of a bilayer system, we need
to find the current densities ${\bf J}_{\mu}$
for given electric
fields ${\bf E}_{\mu}$ in each layer,
where $\mu = \uparrow$ and $\downarrow$
denote the upper and lower layer respectively.
A general configuration $\{ {\bf E}_{\mu}\}$
is a sum of identical and opposite
fields $({\bf E}_{\uparrow}, {\bf E}_{\downarrow})
={\bf E}_{+}(1,1)+{\bf E}_{-}(1,-1)$;
${\bf E}_{\pm} = \frac{1}{2}({\bf E}_{\uparrow}
\pm {\bf E}_{\downarrow})$.
The cases (${\bf E}_{+}\neq 0, {\bf E}_{-}=0$) and
(${\bf E}_{+}=0, {\bf E}_{-}\neq 0$) will be
referred to as the ``in-phase" and ``out-of-phase" mode. Their
corresponding current responses  will be denoted by a subscript ``+"
and
``$-$" respectively.
In linear response, it is sufficient to consider these modes
separately. The ``in-phase" mode is the usual QH transport, with
current response
\begin{equation}
({\bf J}_{\uparrow}+{\bf J}_{\downarrow})_{+}=
\nu\frac{e^{2}}{h}{\bf \hat{B}}\times{\bf E}_{+}, \,\,\,
({\bf J}_{\uparrow}-{\bf J}_{\downarrow})_{+}=0 \label{in} \end{equation}
where ${\bf \hat{B}}$ is the direction of the magnetic field,
taken to be normal to the layers in our discussions.
The ``out-of-phase" mode is unique to bilayer systems. For QHFs, we
shall show that d.c.fields ${\bf E}_{-}$ will generate time
dependent currents.
The sum $({\bf
J}_{\uparrow}+{\bf J}_{\downarrow})_{-}$ oscillates with
period $T=2T_{o}$, where $T_{o}$ is the Josephson period
$e\Delta V/\hbar$, and $\Delta V$ is the voltage that
generates ${\bf E}_{-}$. The factor of 2 is a reflection of
the double-valueness  of the pseudo-spin 1/2.  The time average
current response (denoted by a ``bar") can be determined from general
argument and are given by
\begin{equation}
\overline{({\bf J}_{\uparrow}+{\bf J}_{\downarrow})_{-}}=
n_{A}\nu\frac{e^{2}}{h}{\bf \hat{B}}\times {\bf
E}_{-} , \,\,\,
\overline{({\bf J}_{\uparrow}-{\bf J}_{\downarrow})_{-}}=
n_{B}\frac{\nu}{2}\frac{e^{2}}{h}{\bf \hat{B}}\times {\bf
E}_{-} , \label{out}
\end{equation}
where $n_{A}$ and $n_{B}$ are integers. By studying a number of
examples consistent with the driving force ${\bf E}_{-}$,
we find $n_{A}=0$, $n_{B}=1$.
The sum of eq.(\ref{in}) and (\ref{out}) then gives
\begin{equation}
\overline{ \left( \begin{array}{c} {\bf J}_{\uparrow} \\
{\bf J}_{\downarrow} \end{array} \right) }
 = \nu \frac{e^{2}}{8 h} \left( \begin{array}{cc} 3 & 1 \\1 & 3
\end{array} \right) {\bf \hat{B}}\times
\left( \begin{array}{c} {\bf E}_{\uparrow}
\\ {\bf E}_{\downarrow} \end{array}\right)
\label{result}\end{equation}
These  examples also reveal an interesting oscillatory
behavior of the currents. The time average of $({\bf J}_{\uparrow}+{\bf
J}_{\downarrow})_{-}$ over the half period $T_{o}$ is
$\pm\nu(e^{2}/h){\bf \hat{B}}\times{\bf E}_{-}$,
with alternating sign every
half period. The time average of
$({\bf
J}_{\uparrow}-{\bf J}_{\downarrow})_{-}$ is
$(\nu/2)(e^{2}/h){\bf \hat{B}}\times{\bf E}_{-}$ for all half periods.

To derive these results, we recall that the
Hamiltonian of the bilayer system in the weak tunneling limit
is\cite{MacDonald}
\begin{eqnarray}
H=&\frac{1}{2m}\int d{\bf r}\left[\hat{\psi}^{+}\left({\bf p}-
\frac{e}{c}{\bf A}
\right)^{2}\hat{\psi}-\Delta \hat{m}_{1}({\bf r})\right] \nonumber \\
  +\frac{1}{2}&\int \left[\hat{\rho}({\bf r})\hat{\rho}({\bf r}')
v({\bf r}-{\bf r}')+
\hat{m}_{3}({\bf r})\hat{m}_{3}({\bf
r}')v_{1}({\bf r}-{\bf r}')\right] \label{hamiltonian}
\end{eqnarray}
where ${\bf r}=(x,y)$, $\hat{\psi}_{\mu}({\bf r})$ creates an electron
at the $\mu$ layer at ${\bf r}$,
$\hat{m}_{i}=\frac{1}{2}Tr
\hat{\psi}^{+}\sigma_{i}\hat{\psi}, (i=1,2,3)$, $\hat{\rho}=
Tr\hat{\psi}^{+}\hat{\psi}$
and both $v,v_{1}>0$.
The pseudo-spin axes
${\bf \hat{x}}_{1}, {\bf \hat{x}}_{2}, {\bf
\hat{x}}_{3}$ are chosen so that the symmetric and
antisymmetric state (with energy difference $\Delta$) are
represented by  $\left(^{1}_{1}\right)$ and
$\left(^{1}_{-1}\right)$. They have
no relations with the real space directions ${\bf
\hat{x}}, {\bf
\hat{y}}, {\bf
\hat{z}}$.

We shall consider the
$(mmm)$ state in the correlation limit
$\Delta << e^{2}/\epsilon_{o} \ell \sim e^{2}/\epsilon_{o}
d$, where $d$ is the spacing between layers, and
$\epsilon_{o}$ is the dielectric constant. We begin by expressing their
wavefunctions
(in the symmetric gauge) in spinor form
\begin{equation}
\Psi\left(^{{\bf r}_{1}}_{\mu_{1}}, ...^{{\bf r}_{N}}_{\mu_{N}}\right)
=e^{-\sum_{i=1}^{N}|z|^{2}_{i}/4}\prod_{i=1}^{N}\zeta_{\mu_{i}}
\prod_{i>j}(z_{i}-z_{j})^{m},  \label{wavefunction} \end{equation}
where $z = x+iy$ and
\begin{equation}
\zeta=\left(^{u}_{v}\right)
 = |\zeta|
\left( ^{ e^{-i\phi/2}{\rm cos}\theta/2}_{
e^{i\phi/2}{\rm
   sin}\theta/2} \right)
   e^{-i\chi/2}  \label{spinor}
   \end{equation}
is a spinor along $\hat{\ell}=
{\bf \hat{x}}_{3}{\rm cos}\theta +
{\rm sin}\theta({\bf \hat{x}}_{1}{\rm cos}\phi + {\bf
\hat{x}}_{2}{\rm
sin}\phi)$.
Since $v_{1}>0$, it is energetically more favorable for
$\hat{\ell}$ to lie in the
x$_{1}$-x$_{2}$ plane. $\Delta$ will line up
$\hat{\ell}$ along ${\bf \hat{x}}_{1}$. However, in the correlation limit,
it remains small compared to the correction energy.
The symmetric and antisymmetric states can therefore be regarded as essentially
degenerate $--$ a view that we shall adopt in our subsequent
discussions for simplicity\cite{comment2}.

 If uniform potentials
$V_{\uparrow}$ and $V_{\downarrow}$ are applied to
in the upper and lower layer, $u$ and
$v$ will evolve as $e^{ieV_{\uparrow}t/\hbar}$ and
 $e^{ieV_{\downarrow}t/\hbar}$, $(e>0)$, meaning
\begin{equation}
- \partial_{t}\phi=\frac{e}{\hbar}(V_{\uparrow}-
V_{\downarrow})
\equiv \frac{eV}{\hbar},\,\,\,
- \partial_{t}\chi=\frac{e}{\hbar}(V_{\uparrow}+
V_{\downarrow})\equiv\frac{2eU}{\hbar}. \label{Josephson} \end{equation}
Eq.(\ref{Josephson}) implies means that $\hat{\ell}$
precesses about ${\bf \hat{x}}_{3}$ with angular
frequency $eV/\hbar$.
This immediately implies
an ``unstable" situation if $V$ is
different at two points far apart, (i.e. ${\bf E}_{-}\neq 0)$.
The $\hat{\ell}$ vectors at these points will assume their ground
state configurations, (lying in x$_{1}$-x$_{2}$ plane) but rotate
about ${\bf \hat{x}}_{3}$ at different rates. As a result,
more and more
winding in the spin texture is produced which can only be
relieved by phase slippage processes. As we shall see, during phase
slippage, $\hat{\ell}$ will be pulled away from the
x$_{1}$-x$_{2}$ plane every now and then in some region in space.
While this costs nucleation energy because of the last term in
eq.(\ref{hamiltonian}), it can not be stopped because its
energy cost will evenetually be surpassed by the gradient energy
generated by ${\bf E}_{-}$.

Nonuniform textures implies spatially varying spinors.
However, to stay in the lowest
Landau, $\zeta$ {\em must be analytic in z, (or
 a function of}
$\partial_{z}$).
For simplicity, we shall from now on focus on the ``quasihole" case
where $\zeta(z)$ is an analytic function of $z$.
In general, $\zeta$
can be expanded in powers of $z$.
The simplest example of a nonuniform analytic spinor is
$\zeta(z)
= \alpha + \beta z$, where $\zeta$ reduces to spinors $\alpha$ and $\beta$
as $z$ approaches $0$ and $\infty$.  It also reduces to the usual
quasihole excitations if $\alpha \propto \beta$. For arbitrary
$\zeta$,
eq.(\ref{wavefunction}) implies that
\begin{equation}
2{\bf m}({\bf r}) = \rho({\bf r}) \hat{\ell}({\bf r}), \label{ferro}
\end{equation}
This is a statement of ``full" ferromagnetism and is important  for
our
subsequent discussions.

The relations between density and spin fluctation can be
obtained from the ``plasma analog"\cite{Laughlin}, where one interprets
$\sum_{[\mu]}|\Psi([{\bf r}, \mu])|^{2}$ as the partition function of
a classical 2D plasma with charge $m$ in a background potential
$V_{b}(z) = -|z|^{2}/2 + {\rm ln}|\zeta|^{2}$. This implies that the
electron density is $\rho({\bf r})=$
$=-(4\pi m)^{-1}\nabla^{2}V_{b}$.
$=(2\pi m)^{-1}$$-(4\pi m)^{-1}
{\bf \nabla}^{2}{\rm ln}|\zeta|^{2}$.
Because of the $analyticity$ of $\zeta$, the
density change can be expressed in terms of a deformation field\cite{vs}
${\bf u} = -i\zeta^{+}{\bf \nabla}\zeta/|\zeta|^{2} + c.c.$
\begin{equation}
\rho({\bf r}) = \frac{1}{2\pi m} - \frac{1}{4\pi m}
{\bf \hat{z}}\cdot {\bf \nabla} \times {\bf u} \label{spincharge}
\end{equation}
If follows from eq.(\ref{spinor}) that\cite{MerminHo}\cite{derivation}
\begin{equation}
 {\bf \hat{z}}\cdot {\bf \nabla} \times {\bf u}
= - ({\bf \hat{z}}\times{\bf \nabla})\cdot{\bf \nabla}\chi
+\frac{1}{2}\epsilon_{\alpha\beta\gamma}\ell_{\alpha}
({\bf \hat{z}}\times{\bf \nabla})\ell_{\beta}
\cdot{\bf \nabla}\ell_{\gamma}.
\label{curlvs} \end{equation}
The first term vanishes unless $\chi$ is a vortex, in which
case it is a $\delta$-function\cite{delta}.

The total electric current ${\bf J} = {\bf J}_{\uparrow} + {\bf
J}_{\downarrow}$ can be obtained by differentiating
the number density eq.(\ref{spincharge})
and using the continuity equation
$(-e)\dot{\rho}=-{\bf \nabla}\cdot{\bf J}$,
\begin{equation}
J_{\alpha} = \frac{1}{m}\frac{e^{2}}{h}({\bf \hat{z}}\times
{\bf \nabla})_{\alpha}U -
\frac{e}{4\pi m}\hat{\ell}\cdot
({\bf \hat{z}}\times{\bf
\nabla})_{\alpha}\hat{\ell}\times\partial_{t}\hat{\ell}
\label{chargecur} \end{equation}
where we have made use eq.(\ref{Josephson}). The first term is the usual
QH transport, eq.(\ref{in}). (Note that ${\bf E}_{+}= -{\bf \nabla}U$
and that eq.(\ref{wavefunction}) is derived for ${\bf \hat{B}}= -{\bf
\hat{z}}$).
The second term is due to
the motion of the spin
texture, which is generated by ${\bf E}_{-}$.
To simplify discussions and formulas, we shall from now on focus on
the out-of-phase mode, i.e. ${\bf \nabla}U=0, {\bf \nabla}V\neq 0$. We
shall then drop the first terms
in eq.(\ref{chargecur}) and eq.(\ref{curlvs}).

Consider now the time average of the current in $y$ over a period $T$,
$\overline{(I_{\uparrow} + I_{\downarrow})_{y}}=
T^{-1}\int^{T}_{0}dt\int J_{y}dx$.
The integral is evaluated for an arbitrarily $y$
coordinate, with upper and lower limit $x_{a}$ and $x_{b}$ denoting
the edges of the sample. From eq.(\ref{chargecur}) we have
\begin{equation}
\overline{(I_{\uparrow}+I_{\downarrow})_{y}}=
\frac{- e}{4\pi m}\frac{1}{T}\left[\int^{T}_{0}dt \int^{x_{b}}_{x_{a}} dx
\,\, \hat{\ell}\cdot
\partial_{x}\hat{\ell}\times\partial_{t}\hat{\ell}\right]
\label{oopplus}
\end{equation}
The quantity in the bracket in eq.(\ref{oopplus}) (denoted as $A$) is the area
on an unit sphere $S_{2}$ covered by the ``spactime
rectangle" $\Gamma$ with area $T\times L$, $L=(x_{b}-x_{a})$.
A nonvanishing $A$ means nucleation of vorticity in spacetime
$[\partial_{t}(\nabla\times{\bf u})\neq 0]$.

Eq.(\ref{oopplus}) alone cannot determine
$I_{\uparrow y}$ and $I_{\downarrow y}$.
{}From eqs.(\ref{ferro}),
(\ref{spincharge}), and (\ref{curlvs}), we have
\begin{equation}
2{\bf m}({\bf r}) = \frac{1}{2\pi m}\vec{\ell} - \frac{1}{4\pi
m} \partial_{x}\hat{\ell}\times\partial_{y}\hat{\ell} ,
\label{twospin}
\end{equation}
 \begin{equation}
2\partial_{t}{{\bf m}} = \frac{1}{2\pi m}\partial_{t}\hat{\ell}-
\frac{1}{4\pi
m}\nabla_{\alpha} \left(
({\bf \hat{z}}\times {\bf \nabla})_{\alpha}\hat{\ell}
\times\partial_{t}\hat{\ell}\right) . \label{dmdt}
\end{equation}
Eq.(\ref{dmdt}) says that spin fluctuations will generate
a pseudo-spin current
\begin{equation}
{\bf \vec{K}}_{\alpha} =
\frac{1}{4\pi m}({\bf \hat{z}}\times{\bf \nabla})_{\alpha}
\hat{\ell}\times\partial_{t}\hat{\ell},  \end{equation}
which is related to the charge currents as $(-e){\bf \hat{x}}_{3}\cdot{\bf
\vec{K}}_{\alpha}=J_{\uparrow\alpha}-J_{\downarrow\alpha}$. The time average
of current difference in $y$ is
\begin{equation}
\overline{(I_{\uparrow} - I_{\downarrow})_{y}}=
\frac{- e}{4\pi m}\frac{1}{T}\left[\int^{T}_{0}dt \int^{x_{b}}_{x_{a}} dx
\,\, {\bf \hat{x}}_{3}\cdot
\partial_{x}\hat{\ell}\times\partial_{t}\hat{\ell}\right] .
 \label{B}
\end{equation}
The quantity in the bracket (denoted as $B$) is the area in $S_{2}$ covered
by
$\Gamma$ projected onto the x$_{1}$-x$_{2}$ plane.

To evaluate $A$ and $B$, we consider a rectangular sample (see fig.1)
where the potentials ($V_{\uparrow}, V_{\downarrow}$)
at $x_{a}$ and $x_{b}$ are $(0,0)$ and
$(V/2, -V/2)$ respectively. This is the ``unstable" situation we
discussed before, where $\hat{\ell}$ precesses about
${\bf \hat{x}}_{3}$ with angular frequency $eV/\hbar$ at $x_{b}$
and remains stationary at $x_{a}$\cite{heat}.
In the steady state, the textural motion must be periodic as
${\bf E}_{-}$ forces a constant rotation on $\zeta$ at $x_{b}$.
Note
that $\hat{\ell}$ at $x_{b}$ returns to itself after time $T_{o}=h/eV$,
whereas the period of $\zeta$ (and hence the entire texture
$\hat{\ell}(x,y,t)$) is $T=2T_{o}$.
The factor of 2 is due to the double valueness of $\zeta$.
An example of the steady state evolution of $\hat{\ell}$ is shown in
the spacetime plot $(x$-$t)$ in figure 1. The evolution is a repetition
of the pattern inside the rectangle $T\times L$, ($abb''a''$).
Since the boundary of the rectangle maps onto the equator of $S_{2}$
twice, we then have $A=-4\pi n_{A}$, $B=-2\pi n_{B}$, where
$n_{A}$ and $n_{B}$ are integers depending
on how $S_{2}$ is covered.  Eq.(\ref{oopplus}) and (\ref{B}) then gives
$\overline{(I_{\uparrow}+I_{\downarrow})_{y}}=n_{A}(1/m)(e^{2}/h)(V/2)$,
$\overline{(I_{\uparrow}-I_{\downarrow})}=n_{B}(2m)^{-1}(e^{2}/h)(V/2)$,
which is eq.(\ref{out}) since ${\bf E}_{-}= -\frac{1}{2}{\bf
\nabla}(V_{\uparrow}-V_{\downarrow})=-\frac{1}{2}{\bf \nabla} V$ and
${\bf \hat{B}}=-{\bf
\hat{z}}$.

To determine the topological integers $n_{A}$, $n_{B}$,
we construct explicit examples for
the out-of-phase current response.
Since we are dealing with rectangular geometries, it is more
convenient to use the Landau gauge. The QHF now reads\cite{Thouless}
\begin{equation}
\Psi\left(^{[{\bf r}]}_{[\mu]}\right)
= e^{-\sum_{i}x_{i}^{2}/2}\prod_{i}
\zeta(w_{i})\prod_{i>j}(w_{i}-w_{j})^{m}. \label{Landau}  \end{equation}
where $w=e^{\alpha z}$, $\alpha$ is the ratio of the magnetic
length to the sample length in y, which can be taken as $1$ without
loss of generality. To match the Josephson relations eq.(\ref{Josephson}) at
the boundaries $x_{a}$ and $x_{b}$, (now taken to be $-\infty$ and
$+\infty$), we choose a spinor of the form
\begin{equation}
 \zeta(w) = \lambda \left( \begin{array}{c} 1\\1\end{array}\right)  +
e^{x+iy} \left(\begin{array}{c} e^{ieVt/2\hbar} \\
e^{-ieVt/2\hbar} \end{array} \right) \label{trialspinor} \end{equation}
where $\lambda$ is a constant setting the scale of textural
destortion in real space\cite{last}.
Its textural evolution in the spacetime plane $(x$-$t)$ (for a given
$y$) is shown in fig.1.
$\hat{\ell}$ becomes uniform every half period $T_{o}= h/eV$. The
topological areas $A$ and $B$ for each half peirod can be calculated
from eq.(\ref{trialspinor}) and are independent of $\lambda$.
$A$ is found to be $\mp 2\pi$, with alternating
signs every half period.
$B$ is found to be $-\pi$ for all half periods. This means over the
full period $T=2T_{o}$, $A=0$ and $B=-2\pi$, corresponding to
$n_{A}=0, n_{B}=1$.  We have then derived eq.(\ref{result}) from this
specific example.
I have also examined a number of other
spinors satisfying the same Josephson boundary condition at infinities.
Their time average responses are identical to that of
eq.(\ref{trialspinor}).
I therefore think that these values of $n_{A}$ and
$n_{B}$ have general validity,
despite the lack of a general proof
which is certainly desirable.

This work was stimulated by a talk given by S. Girvin at
the Aspen Center of Theoretical Physics. Various parts of this work
was completed
during a memorable sabbatical visit to HKUST and CUHK in Hong Kong,
and an equally enjoyable visit to NHMFL at Tallahassee.
I thank Nelson Cue, Kenneth Young, Hong-Ming Lai,
and Bob Schrieffer  for their hospitality. This work is supported by
the National High Magnetic Field Laboratory.

\noindent {\bf Figure Caption}

\noindent Figure 1 :
The $\hat{\ell}$ vectors at ${\bf r}$ is
represented by a line emerging from ${\bf r}$. An solid (empty) square
is placed at ${\bf r}$ to indicate that
$\hat{\ell}$ has a positive (negative) $x_{3}$ component.
The boundary of the rectangle $\Gamma=$
$(abb'a')$ maps onto the equator $\bar{\Gamma}$ on the
unit sphere $S_{2}$. The lines $\vec{cc'}$, $\vec{dd'}$, $\vec{ee'}$,
$\vec{c'c''}$, $\vec{d'd''}$, and $\vec{e'e''}$
map onto loops $C,D,E,\bar{C},\bar{D}, \bar{E}$. This covering gives
$A=2\pi$, $B=-\pi$ for the lower rectangle
$abb'a'$ and $A=-2\pi,B=-\pi$ for the upper rectangle $a'b'b''a''$. If the
loops $\bar{C}, \bar{D}$ and $\bar{E}$  are traversed in directions
opposite to those depicted,
then $A=2\pi, B=\pi$ for $a'b'b''a''$.
The textural pattern is calculated from eq.(\ref{trialspinor}) with
$T_{o}=1, y=2, \lambda=8$.


\begin{references}
\bibitem{per} Permanant address.
\bibitem{Girvin} K. Yang, K. Moon, L. Zhang, A.H. MacDonald, S.M.
Girvin, D. Yoshioka, and S.C. Zhang, to be published.
\bibitem{expt} Y.W. Suen, L.W. Engel, M.B. Santos, M. Shayegan, D.C. Tsui,
 Phys. Rev. Lett. {\bf 68}, 1379
(1992); J.P. Eisenstein, G.S. Boebinger, L.N. Pfeiffer, K.W. West, adn Song He,
Phys. Rev. Lett. {\bf 68}, 1383
(1992).
\bibitem{ATT} S.Q. Murphy, J.P. Einstein, G.S. Boebinger, L.N.
Pfeiffer, and K.W. West, to be published.
\bibitem{WenZee} X.G. Wen and A. Zee, Phys. Rev. Lett. {\bf 69}, 1811
(1992);
X.G. Wen and A. Zee, Phys. Rev. B {\bf 47}, 2265 (1993).
\bibitem{comment1} This is because the
``plasma" matrix
$(^{m}_{n}$$^{n}_{m})$ of the $(mmn)$ state vanishes when
$m=n$.
\bibitem{MacDonald} A.H. MacDonald, P.M. Platzman, and G.S. Boebinger,
Phys. Rev. Lett. {\bf 65}, 775, (1990).
\bibitem{comment2} The assumption of degeneracy is not essential
for our results. The derivation including non-vanishing
$\Delta$ is more complicated
and will be discussed elsewhere.
\bibitem{Laughlin} R.B.  Laughlin, Phys. Rev. Lett. {\bf 50}, 1395
(1983).
\bibitem{vs} When expressed in polar angles,
this is exactly the superfluid
velocity ${\bf v}_{s}$
of $^{3}$He-A. See N.D. Mermin and T.L. Ho, Phys. Rev. Lett.
{\bf 36}, 594, (1976), T.L. Ho, Phys. Rev. B, {\bf 18}, 1144 (1978).
However, unlike ${\bf v}_{s}$, ${\bf u}$ is invariant under time
reversal.
\bibitem{MerminHo} This is the vorticity equation of
$^{3}$He-A. (See N.D. Mermin and T.L. Ho, ibid; and T.L. Ho, ibid).
The relation of density and spin fluctuation for polarized QH systems
was pointed out by S.L. Sonhi et.al in Phys. Rev. B {\bf 47}, 16419,
(1993) in the context of Chern-Simon theory. It is also derived by the
authors in ref.4 using projected operator methods. Our derivation here
is different from these approaches. It also reveals some rather
subtle identities which has significant implications on the energetics
of the textural excitations. (See ref.10).
\bibitem{derivation} Eq.(\ref{curlvs}) can be easily obtained
using the identity $\zeta_{\mu}\zeta^{\ast}_{\nu}=\frac{1}{2}
(\delta_{\mu\nu}+\hat{\ell}\cdot\vec{\sigma}_{\mu\nu})|\zeta|^{2}$.
For analytic spinors, (quasihole type), their textures satisfies
$({\bf \hat{z}}\times{\bf \nabla})_{\alpha}\hat{\ell}={\bf
\nabla}_{\alpha}\hat{\ell} \times \hat{\ell}$, which implies
$\nabla_{\alpha}\ell_{\beta}\nabla_{\alpha}\ell_{\beta}=
2\hat{\ell}\cdot\partial_{x}\hat{\ell}\times\partial_{y}\hat{\ell}$.
The integral $\frac{1}{2}\int d{\bf r}
(\nabla_{\alpha}\hat{\ell})^{2}$ is therefore a
topological constant $4\pi n$ if the texture is uniform at infinity.
\bibitem{delta} A singular vortex can be generated if $\zeta$ contains
an usual quasihole, i.e. both $u$ and $v$ have a common factor
$(z-a)$.
Note that the plasma analog is only
accurate on length scales larger than the magnetic length. All
$\delta$-functions revealled by the plasma analog are in fact
Guassians.
\bibitem{heat} The situation here
is identical to $^{3}$He-A in a heat flow, where
the texture is continously being wounded up by a
chemical potential gradient. The winiding is undone by a periodic
nucleation of coreless vortices.
which is relieved by a periodic texture motion. See
T.L. Ho, Phys. Rev. Lett. {\bf 41}, 1473 (1978).
\bibitem{Thouless} Eq.(\ref{ferro}) to (\ref{curlvs}) can also be
derived from eq.(\ref{Landau}).
The plasma analog can be generalized easily to the Landau gauge by
realizing that density of the plasma $\rho(w)$ is related to the
electron density $\rho(z)$ by the Jacobian $|dw/dz|^{2}$.
\bibitem{last} It follows from eq.(\ref{twospin}) that for
spinor eq.(\ref{trialspinor}), $\int m_{3}=0$. This is consistent with
the degenerate (or zero tunneling)
limit $\Delta\rightarrow 0$ that we have taken.
\end{references}
\end{document}